\documentstyle[12pt,mysymbols]{article}
\input psbox.tex

\newcommand{\be}{\begin{equation}}
\newcommand{\ee}{\end{equation}}
\newcommand{\bea}{\begin{eqnarray}}
\newcommand{\eea}{\end{eqnarray}}
\newcommand{\brr}{\begin{array}}
\newcommand{\err}{\end{array}}
\newcommand{\bc}{\begin{center}}
\newcommand{\ec}{\end{center}}
\newcommand{\nn}{\nonumber}
\newcommand{\sss}{\scriptscriptstyle}

\newcommand{\GF}{\frac{G_{\sss F}}{\sqrt 2}}

\newcommand{\DSone}{\Delta S\!=\!1}

\newcommand{\ms}{m_{\sss s}}
\newcommand{\mc}{m_{\sss c}}
\newcommand{\mb}{m_{\sss b}}
\newcommand{\mt}{m_{\sss t}}
\newcommand{\Mw}{M_{\sss W}}

\newcommand{\W}{\hat{\bb W}}

\newcommand{\one}{O_{\sss 1}}
\newcommand{\two}{O_{\sss 2}}
\newcommand{\onec}{O_{\sss 1}^{\sss c}}
\newcommand{\twoc}{O_{\sss 2}^{\sss c}}

\newcommand{\six}{O_{\sss 6}}

\newcommand{\threefive}{O_{\sss 3,5}}
\newcommand{\foursix}{O_{\sss 4,6}}
\newcommand{\sevennine}{O_{\sss 7,9}}
\newcommand{\eightten}{O_{\sss 8,10}}
\newcommand{\alphas}{\alpha_{\sss s}}
\newcommand{\alphae}{\alpha_{\sss e}}
\newcommand{\alphass}{\alpha_{\sss s}^{\sss 2}}
\newcommand{\gammaszero}{\hat\gamma_{\sss s}^{\sss (0)}}
\newcommand{\gammasone}{\hat\gamma_{\sss s}^{\sss (1)}}
\newcommand{\gammaezero}{\hat\gamma_{\sss e}^{\sss (0)}}
\newcommand{\gammaeone}{\hat\gamma_{\sss e}^{\sss (1)}}
\newcommand{\Otwoc}{\Omega_2^c}
\newcommand{\Ofour}{\Omega_4}
\newcommand{\Oseven}{\Omega_7^{\sss 3/2}}
\newcommand{\Oeight}{\Omega_8^{\sss 3/2}}
\newcommand{\Onine}{\Omega_9^{\sss 3/2}}
\newcommand{\Osen}{\Omega_{7,8,9}^{\sss 3/2}}
\newfont{\bb}{msbm10   scaled\magstep1}
\begin{document}
\begin{flushright}
ULB-TH 93/04\\
April 1993\\
\end{flushright}

\vskip 3.5 truecm
\centerline{\bf{NLO CALCULATION OF $\epsilon^{\prime}/\epsilon$ IN
QCD AND QED}}
\centerline{\bf{WITHIN THE LATTICE QCD FRAMEWORK.}}
\vskip 1. truecm
\centerline{L. Reina}
\centerline{Service de Physique Th\'eorique,
Universit\'e Libre de Bruxelles,}
\centerline{Campus Plaine - CP 225, Boulevard du Triomphe}
\centerline{B-1050 Brussels, Belgium. }
\vskip 5.5 truecm
\begin{abstract}
We present\footnote{Work done in collaboration with M. Ciuchini,
E. Franco and G. Martinelli}
the results of our recent calculation of $\epsilon^{\prime}/
\epsilon$ at the Next-to-Leading (NLO) order in QCD and QED, in the
framework of $\DSone$ Effective Hamiltonians. The operator matrix
elements are taken from lattice QCD, at a scale $\mu=2\,\,\mbox{GeV}$.
NLO corrections seem to lower the value of
$\epsilon^{\prime}/\epsilon$, favouring the experimental result of
E731. Contributions from different operators are analyzed and
dependence on $\mt$, $\mu$ and $\Lambda_{\sss QCD}$ discussed.
\end{abstract}
\vskip 5mm
\newpage

\subsection*{Introduction}

Quark mixing and CP-violation are two important subjects of the
Standard Model physics and might suggest the presence of new physics
beyond it.  Among the reasons which might justify the interest in the
physics of CP-violation and flavour mixing, the possibility to probe
the perturbative and non-perturbative theoretical consistence of the
model and to investigate the physics of heavy flavours are maybe the
most important ones.

$K$-meson and $B$-meson systems are the natural framework where the
physics of quark mixing and CP-violation is studied. Flavour-mixing
and CP-violating weak decays are described by Effective Hamiltonians
of the form:
\be
{\cal H}_{\sss eff}=\sum_{\sss i}C_{\sss i}(\mu)O_{\sss i}(\mu)
\ee
where the product of currents present in the true hamiltonian is
expanded through the OPE (Operator Product Expansion), at a scale
$\mu$, as a sum over a given set of operators
$\{O_{\sss i}(\mu)\}$ with suitable Wilson coefficients
$\{C_{\sss i}(\mu)\}$. The short and long distance
parts of the problem factorize into the Wilson coefficients and the
operators respectively. In the evaluation of physical amplitudes, the
Wilson coefficients are treated perturbatively, while the matrix elements
of the effective operators $\{O_{\sss i}\}$ are derived with
non-perturbative techniques.

In particular we have calculated the Effective Hamiltonian for
$\DSone$ decays at NLO in QCD and QED \cite{cfmr1,cfmr2}. The Wilson
coefficients of the $\DSone$ Effective Hamiltonian have been derived
from the ($10\times 10$) anomalous dimension matrix, which includes
corrections at orders $\alphas$, $\alphass$, $\alphae$,
$\alphas\alphae$. The matrix elements for the corresponding operators
have been taken from lattice QCD .

The knowledege of the $\DSone$ Effective Hamiltonian at NLO in QCD and
QED is important for several reasons:
\begin{itemize}
\item heavy mass (like $\mt$) effects are indeed a NLO effect;
\item the stability of the perturbative calculation can be checked
and a limit on its reliability can be fixed;
\item the $\Lambda_{\sss QCD}$ parameter can be used in a proper way,
taking it from different experiments;
\item the effect of different operators in the OPE (QCD-penguins,
QED-penguins, etc.) can be better determined.
\end{itemize}

We finally present our results for the $\epsilon^{\prime}/\epsilon$
parameter of direct CP-violation in $K$-decays. The effect of QCD and
QED corrections at NLO seems to lower the value of $\epsilon^{\prime}/
\epsilon$, favouring the experimental value of E731 \cite{e731}.

\subsection*{$\DSone$ Effective Hamiltonians at NLO: general results}

The Effective Hamiltonian for $\DSone$ decays is given by:
\bea
{\cal H}_{eff}^{\DSone}&=&\lambda_{\sss u}\GF
\Bigl[ (1 - \tau ) \Bigl( C_{\sss 1}(\mu)\left(\one(\mu) -
\onec(\mu) \right) + C_{\sss 2}(\mu)\left( \two(\mu) - \twoc(\mu)
\right)  \Bigr)\nn\\
&+&\tau\sum_{\sss i=3,\ldots ,10} O_{\sss i}(\mu) C_{\sss i}(\mu)
\Bigr]
\label{eh}
\eea
where $\lambda_{\sss u} = V_{\sss ud} V^*_{\sss us}$ and similarly we can
define $\lambda_{\sss c}$ and $\lambda_{\sss t}$. $\tau=-\lambda_{\sss t}/
\lambda_{\sss u}$ and $V_{\sss ij}$ is one of the elements of the
CKM mixing matrix.

We have used the following complete basis of operators when QCD and QED
corrections are taken into account:
\bea
\mbox{Vertex-type}&\rightarrow&
\left\{\begin{array}{l}
\one = ({\bar s}_{\sss\alpha}d_{\sss\alpha})_{\sss (V-A)}
      ({\bar u}_{\sss\beta}u_{\sss\beta})_{\sss (V-A)} \\
\two = ({\bar s}_{\sss\alpha}d_{\sss\beta})_{\sss (V-A)}
      ({\bar u}_{\sss\beta}u_{\sss\alpha})_{\sss (V-A)} \\
\onec = ({\bar s}_{\sss\alpha}d_{\sss\alpha})_{\sss (V-A)}
      ({\bar c}_{\sss\beta}c_{\sss\beta})_{\sss (V-A)} \\
\twoc = ({\bar s}_{\sss\alpha}d_{\sss\beta})_{\sss (V-A)}
      ({\bar c}_{\sss\beta}c_{\sss\alpha})_{\sss (V-A)}
\end{array}\right. \nn\\
\mbox{QCD-Penguins}&\rightarrow&
\left\{\begin{array}{l}
\threefive = ({\bar s}_{\sss\alpha}d_{\sss\alpha})_{\sss (V-A)}
       \sum_{\sss q=u,d,s}({\bar q}_{\sss\beta}q_{\sss\beta})_{\sss (V\mp A)}\\
\foursix = ({\bar s}_{\sss\alpha}d_{\sss\beta})_{\sss (V-A)}\sum_{\sss q=u,d,s}
      ({\bar q}_{\sss\beta}q_{\sss\alpha})_{\sss (V\mp A)}
\end{array}\right.\nn\\
\mbox{QED-Penguins}&\rightarrow&
\left\{\begin{array}{l}
\sevennine = \frac{3}{2}({\bar s}_{\sss\alpha}d_{\sss\alpha})_{\sss (V-A)}
      \sum_{\sss q=u,d,s} e_{\sss q}({\bar q}_{\sss\beta}q_{\sss\beta})
       _{\sss (V\pm A)} \\
\eightten = \frac{3}{2}({\bar s}_{\sss\alpha}d_{\sss\beta})_{\sss (V-A)}\sum_q
      e_{\sss q}({\bar q}_{\sss\beta}q_{\sss\alpha})_{\sss (V\pm A)}
\end{array}\right.
\label{epsilonprime_basis}
\eea
where the subscript $(V \pm A)$ indicates the chiral structure and
$\alpha$ and $\beta$ are colour indices.
\par The operators $O_{\sss i}(\mu)$ are renormalized at the scale
$\mu<\Mw$ in $\overline{MS}$ in some given regularization scheme (e.g.
HV ('t Hooft-Veltman) or NDR (Naive Dimensional Reduction)). The
corresponding coefficients, $C_{\sss i}$, are also scheme dependent.

We have matched the full and the effective theory at $\mu=\Mw$. This
gives the initial conditions for the evolution of the coefficient
functions. It's here that the main dependence on the heavy top mass
enters. Then, at a generic scale $\mu<\Mw$, the NLO evolved coefficient
function can be expressed as \cite{cfmr1,cfmr2,bur4,bjlw,bjl}:
\be
\vec C(\mu) = \W[\mu,\Mw] \vec C(\Mw)
\label{evo}
\ee
where $\vec C(\mu)$ is a vector, whose components are the
corresponding Wilson coefficients. $\vec C(\Mw)$ is the vector of the
initial conditions at $\mu=\Mw$ and $\W[\mu,\Mw]$ the renormalization
group evolution matrix. The matrix $\W[\mu,\Mw]$ depends on the
one-loop and two-loop coefficients of the Anomalous Dimension Matrix
(ADM) $\hat\gamma$ for the operator basis in (\ref{epsilonprime_basis}):
\be
\hat \gamma= \frac {\alphas }{ 4 \pi } \gammaszero +
\frac {\alphae }{4 \pi} \gammaezero
+ \frac {\alphas^{\sss 2} }{(4 \pi)^{\sss 2}} \gammasone +
\frac{ \alphas }{4 \pi} \frac{ \alphae}{4 \pi}  \gammaeone \nn
\ee
Both $\vec C(\Mw)$ and $\W[\mu,\Mw]$ are regularization scheme
dependent. We have computed $\vec C(\Mw)$, $\gammasone$ and
$\gammaeone$ both in HV and NDR scheme and verified the scheme
independence of the final result (\ref{eh}). Moreover, we have
checked our results both at the matrix level and at the diagram by
diagram level.
\par
In ref.\cite{cfmr2} we discuss all the technical details of our
calculation and the differences between our results and those obtained
in ref.\cite{bur4,bjlw,bjl} by the Munich group.

\subsection*{Results for $\epsilon^{\prime}/\epsilon$}

In the expression for $\epsilon^{\prime}$ ($e^{{\mbox \tiny i}\delta_{\sss i}}
A_{\sss i}=\langle \pi\pi(I=i)|{\cal H}_{eff}^{|\Delta S|=1}|K\rangle $):
\be
\epsilon^{\prime}=\frac{e^{\sss i\pi/4}}{\sqrt{2}}\frac{\omega}
{\mbox{Re}A_{0}}\left[\omega^{ -1}
(\mbox{Im}A_{2})^{\prime}-(1-\Omega_{\sss IB})\,\mbox{Im}A_{0}
\right]
\label{epsilonprime}
\ee
the real parts $\mbox{Re}A_{0}$ and $\mbox{Re}A_{2}$ are taken from
experiments ($\omega=\mbox{Re}A_{0}/\mbox{Re} A_{2}=0.045$), while
the imaginary parts
$\mbox{Im}A_{0}$ and $(\mbox{Im}A_{2})^{\prime}$ can be derived from
${\cal H}_{eff} ^{\DSone}$ in the following form:
\bea
\mbox{Im}A_{ 0} &=&-\GF\mbox{Im}\left({\bb V}_{ts}^{*}{\bb V}_{td}\right)
\left\{-\left(C_{ 6}B_{ 6}+\frac{1}{3}C_{ 5}B_{ 5}\right)Z
+\left(C_{ 4}B_{ 4}+\frac{1}{3}C_{ 3}B_{ 3}\right)X+\right.
\nn\\
& &\!C_{ 7}B_{ 7}^{ 1/2}\left(\frac{2Y}{3}+\frac{Z}{6}-
\frac{X}{2}\right)+C_{ 8}B_{ 8}^{ 1/2}\left(2Y+\frac{Z}{2}+
\frac{X}{6}\right)-\nn\\
& &\!\left.C_{ 9}B_{ 9}^{ 1/2}\frac{X}{3}+\left(\frac{C_{ 1}
B_{ 1}^{ c}}{3}+C_{ 2}B_{ 2}^{ c}\right)X\right\}
\label{ima0}
\eea
and
\bea
(\mbox{Im}A_{ 2})^{\prime}\!&=&\!-\GF\mbox{Im}\left({\bb V}_{
ts}^{*}{\bb V}_{ td}\right)
\left\{C_{ 7}B_{ 7}^{ 3/2}\left(\frac{Y}{3}-\frac{X}{2}\right)+
\right. \nn \\
& &\!\left.C_{ 8}B_{ 8}^{ 3/2}\left(Y-\frac{X}{6}\right)+
C_{ 9}B_{ 9}^{ 3/2}\frac{2X}{3}\right\}
\label{ima2}
\eea
where we have introduced $(\mbox{Im}A_{ 2})^{\prime}$ defined as:
\be
\mbox{Im}A_{ 2}=(\mbox{Im}A_{ 2})^{\prime}+\Omega_{\sss IB}
(\omega\,\mbox{Im}A_{ 0})
\ee
$\Omega_{\sss IB}\!=\! 0.25 \pm 0.10$ represents the isospin breaking
contribution, see for example ref. \cite{bur5}.
\par
The Wilson coefficients have been calculated at the NLO in QCD and QED
as explained before. The matrix elements of the corrisponding
operators have been expressed in terms of quantities $X$, $Y$ and $Z$
(see \cite{lmmr,cfmr1}) and B-parameters taken (whenever possible)
from lattice QCD. We recall that the B-parameter for a given operator
is defined as the ratio of its matrix element to the same matrix
element evaluated in the VIA (Vacuum Insertion Approximation). The
values of the still missing B-parameters are guessed on the basis of
some reasonable arguments. In particular, $O_{\sss 7,8,9}^{\sss 1/2}$
turn out to have negligible coefficients; while $O_{\sss 1,2}^{\sss
c}$ and $O_{\sss 3,4}$ have very large coefficients. Thus we have
fixed $B_{\sss 7,8,9}^{\sss 1/2}$ to their VIA value ($=1$); while we have
allowed $B_{\sss 1,2}^{\sss c}$ and $B_{\sss 3,4}$ to vary in a quite
large range around their VIA value ($0$ and $1$ respectively). For a
more detailed discussion and full references on recent lattice
calculations see ref. \cite{lmmr,cfmr1}. We report the values used for
the B-parameters in Table \ref{Bpar}.
\begin{table}
\begin{center}
\begin{tabular}{c c c c c c}\hline\hline\\
$B_{ K},B_{ 9}^{ (3/2)}$ &  $B_{ 1-2}^{ c}$ & $B_{ 3,4}$ &
$B_{ 5,6}$ & $B_{ 7-8-9}^{ (1/2)}$ & $B_{ 7-8}^{ (3/2)}$
\\ \\\hline \\
$0.8\pm 0.2$ &  $0 - 0.15^{ (*)}$ & $1 -  6^{ (*)}$ &
$1.0\pm 0.2$ & $1^{ (*)}$ & $1.0\pm0.2$
\\ \hline\hline
\end{tabular}
\caption[]{Values of the $B$-parameters. Entries with a $^{ (*)}$
are educated guesses; the others are taken from lattice QCD calculations.}
\label{Bpar}
\end{center}
\end{table}

Writing $\epsilon^{\prime}/\epsilon$ in terms of {\it relative}
contributions of different operators with respect to the $\six$
penguin operator, i.e.:
\be
\epsilon^{\prime}/\epsilon  \sim  R   \times
 C_{ 6} B_{ 6} \Bigl( 1- \sum_i \Omega_{ i} \Bigr)
\label{kfact}
\ee
where $\Omega_i=C_i B_i/C_6 B_6$, we get that terms $\Otwoc$, $\Ofour$,
$\Osen$ give the main contributions. A detailed discussion of the
phenomenological analysis performed is presented in ref. \cite{cfmr1}.

Our results can be summarized as follows:
\begin{itemize}
\item Fixing $\mt=140\,\,\mbox{GeV}$, $\mu=2\,\,\mbox{GeV}$ and allowing
B-parameters, $\ms$, $\Lambda_{\sss QCD}$, $\Omega_{\sss IB}$, CKM
parameters, etc. to vary around their central values (see Table
\ref{tab_val}), we get an idea of the theoretical incertainty and of
the influence of NLO corrections. The main observation is that, the
sums $\Otwoc+\Ofour$ and $\Oseven+\Oeight+\Onine$ (despite significant
individual variations from LO to NLO in the last case) are almost {\it
stable} with respect to NLO corrections. Therefore the behaviour of
the central value of $\epsilon^{\prime}/\epsilon$ is still governed by
the contribution of $\six$ and the decreasing of $C_6$ with NLO
corrections lowers the central value of $\epsilon^{\prime}/\epsilon$,
favouring the E731 result (see Figs.(\ref{fig2})-(\ref{fig3})).
\begin{table}
\begin{center}
\begin{tabular}{c c}\hline
{\it parameter} & {\it
value} \\ \hline \\
$\Lambda_{ QCD}$ & $340\pm 120$ GeV \cite{alto,german}\\
$\ms(2\,\mbox{GeV})$ & $(170\pm 30)$ MeV\\
$\mc(2\,\mbox{GeV})$ & $1.5$ GeV \\
$\mb(2\,\mbox{GeV})$ & $4.5$ GeV \\
$A\lambda^2$ & $0.047\pm 0.004$\\
$\sqrt{\rho^2+\eta^2}={ V}_{ ub}/(\lambda{ V}_{ cb})$ & $0.50\pm 0.14$\\
$\epsilon_{exp}$ & $2.28\cdot 10^{-3}$\\
$\mbox{Re}A_{ 0}$ & $2.7\cdot10^{-7}\mbox{GeV}$\\
\hline
\end{tabular}
\caption[]{Values of experimental parameters used in this work.}
\label{tab_val}
\end{center}
\end{table}
\item Varying $\mt$ between $100\,\,\mbox{GeV}$ and $200\,\,\mbox{GeV}$, we
find that NLO corrections are much more important for higher values of
$\mt$. Thus the central value of $\epsilon^{\prime}/\epsilon$ decreases with
increasing $\mt$ (see Figs.((\ref{fig3})-(\ref{fig5})).
\item Varying $\mu$ and $\Lambda_{\sss QCD}$ (see Table \ref{tab_val}), we
note that below $\mu\sim 1\,\,\mbox{GeV}$ the perturbative approach is
not reliable anymore; while the behaviour of the Wilson coefficients
is quite stable for higher values of $\mu$ (see Fig.(\ref{fig1})).
This observation greatly supports the use of lattice results, which
allows to match matrix elements and Wilson coefficients at a scale
$\mu\sim 2\,\,\mbox{GeV}$.
\end{itemize}

\begin{figure}
    \begin{center}
       \mbox{\psbox{ncf68.eps}}
       \caption[]{ $C_6$ and $C_8$ as a function of $\mu$  for
       $\Lambda_{QCD}=220$ (dotted),$ 340$ (solid) and $460$ (dashed) MeV.}
       \label{fig1}
    \end{center}
\end{figure}
\begin{figure}
    \begin{center}
          \mbox{\psbox{nfig2_nlo100_r.eps}}
           \caption[]{ Band of allowed values for $\epsilon^{\prime}/\epsilon$
            at $\mt=100\,\,\mbox{GeV}$ (NLO). The dashed lines represent the
            experimental results of NA31, $(2.3 \pm 0.7)\cdot 10^{-3}$
            \cite{na31} and E731, $(0.74 \pm 0.59)\cdot 10^{-3}$ \cite{e731}.}
           \label{fig2}
    \end{center}
\end{figure}
\begin{figure}
    \begin{center}
          \mbox{\psbox{nfig2_llo100_r.eps}}
          \caption[]{ Same as in Fig.(\ref{fig2}) at  $\mt=100\,\,\mbox{GeV}$,
           with the coefficients of the operators for $\epsilon^{\prime}
           /\epsilon$ computed at the LO.}
          \label{fig3}
    \end{center}
\end{figure}
\begin{figure}
    \begin{center}
          \mbox{\psbox{nfig2_nlo140_r.eps}}
          \caption[]{Same as in Fig.(\ref{fig2}) at  $\mt=140\,\,\mbox{GeV}$.}
          \label{fig4}
    \end{center}
\end{figure}
\begin{figure}
    \begin{center}
          \mbox{\psbox{nfig2_nlo200_r.eps}}
          \caption[]{Same as in Fig.(\ref{fig2}) at  $\mt=200\,\,\mbox{GeV}$.}
          \label{fig5}
    \end{center}
\end{figure}

\end{document}